\documentclass[11pt,twocolumn,a4paper]{article}

\usepackage{bm,amsmath,amssymb,amsfonts}

\usepackage[dvips]{graphicx}
\usepackage[margin=12pt,font=normalsize,labelfont=bf]{caption}

\usepackage{color}
\definecolor{dred}{rgb}{0.8,0,0}
\definecolor{dgreen}{rgb}{0,0.5,0}

\usepackage{setspace}
\begin{document}

\onecolumn

\begin{center}
{\bf{\Large\textcolor{dred}{Electron states and magneto-transport 
in a graphene geometry 
with a fractal distribution of holes}}}\\
~\\
\textcolor{dgreen}{Biplab Pal$^{\textcolor{black}
{1,}}$}\renewcommand{\thefootnote}
{\ensuremath{\dag}}\footnote{{\bf Corresponding Author}: 
Biplab Pal\\
$~$\hspace {0.45cm} Electronic mail: 
biplabpal@klyuniv.ac.in}\textcolor{dgreen}
{, Arunava Chakrabarti$^{\textcolor{black}{1}}$} 
\textcolor{dgreen}{and Nitai Bhattacharya$^{\textcolor{black}
{2}}$}\\
\it $^1$Department of Physics,
University of Kalyani,
Kalyani, West Bengal 741 235, India\\
\it $^2$Nimtala Rangeswar High School, Nadia, West Bengal, India\\
~\\
{\bf Abstract}
\end{center}
We consider an infinite graphene geometry where bonds and sites have been 
removed selectively to map it onto an effective Sierpinski gasket 
comprising of hexagons. We show that such a structure is capable of 
sustaining an infinite number of extended single particle states inspite 
of the absence of any translational order. When each basic hexagonal 
plaquette in the Sierpinski geometry is threaded by a magnetic flux, the 
spectrum exhibits bands of extended eigenstates. The bands persist for 
any arbitrary value of the flux but disappear again as the flux becomes 
equal to half the fundamental flux quantum. The localization - de-localization issues are discussed thoroughly along with the computation 
of two terminal magneto-transport of finite versions of the lattice. The 
numerical results corroborate our analytical findings.
\vskip 0.4cm
\begin{flushleft}
{\bf PACS No.}: 73.43.-f, 73.23.-b, 61.43.Hv, 73.20.At\\
~\\
{\textbf {\itshape Keywords}}: Graphene, Fractal Lattice, Localization, 
Renormalization Group
\end{flushleft}

\newpage
\section*{1. Introduction}

Graphene systems have been at the center stage of research in low 
dimensional systems, in particular, nano-structures, in the last decade. 
Both graphene systems in 2D and graphene nano ribbons have been proposed 
as potential candidates for various applications in 
nanotechnology~\cite{novoselov}~-~\cite{yamashiro}. A wide variety of 
problems have been investigated not only from the angle of applications, 
but also from the standpoint of the novel physics offered by such 
systems. Electronic and magnetic properties of triangular graphene 
rings~\cite{potasz}, structural, mechanical and electrical properties of 
defect-patterned graphene nano-meshes~\cite{sahin}, molecular transistors 
connected to graphene nano-ribbon electrodes~\cite{kamal} are among some 
of the recent exciting studies in the graphene system.

In recent years, several interesting experiments and theoretical works 
have been reported which deal with vacancies in a graphene sheet or 
nanoribbon. Such vacancies are usually created by ion bombardment, and in 
principle, the locations can be controlled almost at will. Pereira and 
Schulz~\cite{schulz} have studied the effects of vacancies on the 
electronic properties of a graphene sheet in the presence of a 
perpendicular magnetic field. The effect of an extended linear defect on 
the electronic transport properties of a graphene sheet has been 
investigated by Bahamon, Pereira and Schulz~\cite{bahamon}. It is shown 
that such defects profoundly modify the properties of a nano-ribbon and 
even introduce new conductance quantization values. Using a first 
principles method Jippo, Ohfuchi and Kaneta~\cite{jippo} report the 
transport properties of graphene sheets having two- and one-dimensional 
periodic nano holes. 

Vacancies in graphene lattice are shown to give rise to {\it extra} 
localized states between Landau levels~\cite{schulz}. The distance 
between the vacancies and their pattern of distribution have non-trivial 
effect on the physical properties of a 2D graphene geometry. The 
modification of the bands, the nature of the wave functions in the 
presence of the vacancies are inspiring in introducing such vacancies on 
purpose following a tailor made design~\cite{schulz}, and look for any 
unusual properties that might suggest new applications.
 
It is thus intriguing to know what happens to the electronic states of a 
graphene sheet when one generates holes of increasing size, and 
distributes them over an infinite two dimensional graphene sheet 
following a pre-determined geometry. This is the central motivation 
behind the present work. We undertake a detailed investigation of the 
single particle states on an infinite  graphene sheet where holes with 
increasing sizes have been distributed following a Sierpinski 
fractal geometry~\cite{rammal}~-~\cite{banavar}. 

Such a study serves a dual purpose. Sierpinski gasket (SPG) fractals have 
been well known candidates for studying percolation clusters. These self 
similar lattices are known to give rise to exotic electronic energy 
spectrum, revealing a fragmented Cantor set character~\cite{rammal}~-~\cite{banavar}, that has been shown later to contain an infinite number 
of extended eigenstates embedded in it~\cite{wang1}~-~\cite{wang2}. 
Recently, a hexagonal Sierpinski structure has been experimentally 
synthesized~\cite{newkome}. The graphene with a Sierpinski distribution 
of holes lies very close to the synthesized structure, and gives us an 
opportunity to analyze the character of electronic states in these 
systems. Secondly, the present day lithographic techniques allows us to 
fabricate networks following any desired geometry. The potential of a 
graphene-fractal in the field of nano devices is thus worth 
investigating.

We work within a single band tight binding formalism. Several works 
within the tight binding formalism and with spinless, non-interacting 
electrons have been successful in highlighting the physical properties of 
graphene systems with or without defects~\cite{schulz}~-~\cite{jippo}. We 
use the same, together with the real space renormalization (RSRG) group 
decimation scheme that exploits the self-similarity of a fractal 
geometry, to unravel the character of single particle eigenstates. 
The influence of a magnetic flux piercing a selected subset of 
the graphene hexagons on its electronic properties, is studied 
in details. 
In 
addition to this, the two terminal electronic transport across 
finite but arbitrarily large graphene-fractals is worked out.

Our results are interesting. We find that in the absence of any magnetic 
field, the graphene-fractal gives rise to an infinite number of extended 
eigenstates which coexist with the fragmented spectrum of localized 
states. The energies of the extended states can be precisely determined 
from the RSRG recursion relations. As soon as a magnetic field is 
switched `on' in an elementary hexagon of the graphene-fractal, the 
density of states exhibits continuous distribution of extended states in 
the spectrum. These {\it bands} of extended states persist for all non-
zero values of the magnetic flux $\Phi$, except for $\Phi = \Phi_0/2$, 
where $\Phi_0 = hc/e$ is the flux quantum. At $\Phi = \Phi_0/2$ the 
spectrum swings back to the flux free shape with sharply localized 
eigenstates dominating the entire spectrum. Incidentally, appearance of 
such bands of extended states has been suggested earlier, based on 
extensive numerical studies by Chakrabarti~\cite{arunava2} and Schwalm \& 
Moritz~\cite{schwalm3} in the context of a $3$-simplex fractal and a 
modified rectangle lattice respectively. These are self similar lattices 
without any translational invariance. The existence of flux driven bands 
of extended states is thus curious. With the results presented in this 
communication we have a strong case where we can conjecture that bands of 
extended states might be a generic feature of deterministic fractals with 
holes. 

In what follows, we describe the results of our calculation. Section 2 
describes the model and the method. In section 3 we extract and analyze 
the extended eigenstates, discuss the density of states with and 
without the magnetic flux threading the basic plaquettes and, 
present the detailed results of the magneto-transport calculations for 
finite graphene-fractals . In section 4 we draw the conclusions. 
\section*{2. The model and the method}

\begin{figure}[ht]
\centering 
\includegraphics[clip,width=7.5cm]{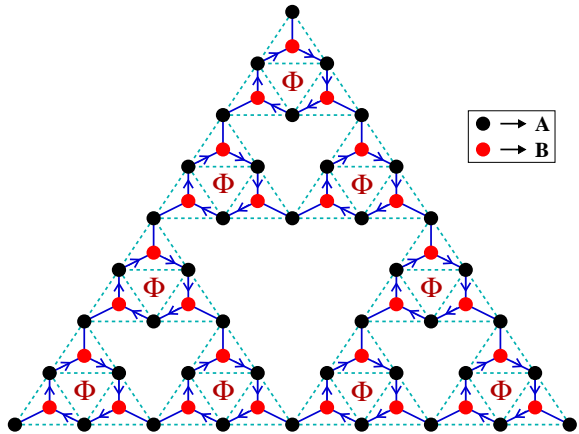}
\caption{Schematic diagram of a portion of the graphene-fractal scooped 
out of an infinite graphene sheet. Each of the surviving hexagonal 
plaquettes is threaded by a uniform magnetic flux $\Phi$. The underlying 
dotted line shows the Seirpinski gasket fractal geometry in which the basic 
graphene-fractal geometry can be mapped onto by decimating the `B' type 
(red) atomic sites.}
\label{lattice1}
\end{figure}

Let us refer to Fig.~\ref{lattice1}. A selective removal of sites from 
the two dimensional graphene sheet will lead to a set of extended defects 
(voids) which distribute themselves on an effective Sierpinski fractal 
network~\cite{rammal}. With both the A (black) and B (red) sites now 
present in the lattice, the two dimensional geometry in fact resembles a 
Zachariascn fractal glass~\cite{schwalm2}. The fluctuating environment 
around each site generates unusual eigenstates and strange transport 
properties as studied earlier~\cite{wang1}~-~\cite{wang2}. We shall be 
concerned with the effect of the parent graphene geometry on the 
electronic states of such a system with multiple holes. To test the 
effect of a magnetic flux on the spectral properties within a minimal 
model, we include a uniform magnetic flux $\Phi$ through each of the 
hexagons in Fig.~\ref{lattice1}. The system is described by the standard 
tight binding Hamiltonian, 
\begin{equation}
H=\sum_{i}\epsilon_{i}|i\rangle\langle{i}|
+\sum_{\langle ij \rangle} \:[\;t_{ij} e^{i \theta_{ij}}|i \rangle 
\langle{j}| + t_{ji} e^{-i \theta_{ij}}|j \rangle \langle{i}|\;]
\label{Hamiltonian}
\end{equation}
where, $\epsilon_i$ is the on-site potential at the $i$-th atomic site 
and $t_{ij}$ is the nearest neighbor hopping integral. $\theta_{ij} = 2 
\pi \Phi/6 \Phi_0$ is the Peierl's phase which is included in the hopping 
along each arm of a hexagon. $\Phi_0 = hc/e$ is the fundamental flux 
quantum. With the selective inclusion of the magnetic flux, the graphene-
fractal now describes a system where the time reversal symmetry is broken 
only on a subset of the bonds. Consequent changes in the energy spectrum 
and transport characteristics thus need to be examined in details. It may 
be mentioned that, the triangular symmetry underlying a graphene topology 
has been exploited previously by Vargas and Naumis~\cite{naumis} to 
explain an increase in localization in a doped graphene. The present 
system is thus worth examining.

For our purpose we shall consider constant values of $\epsilon_i = 
\epsilon_0$ and $t_{ij} = t_0$ throughout the calculation. That is, the 
A- and the B-sites are not distinguished energetically. In the context of 
the graphene sheet they are all carbon atoms.

\begin{figure}[ht]
\centering 
\includegraphics[clip,width=15cm]{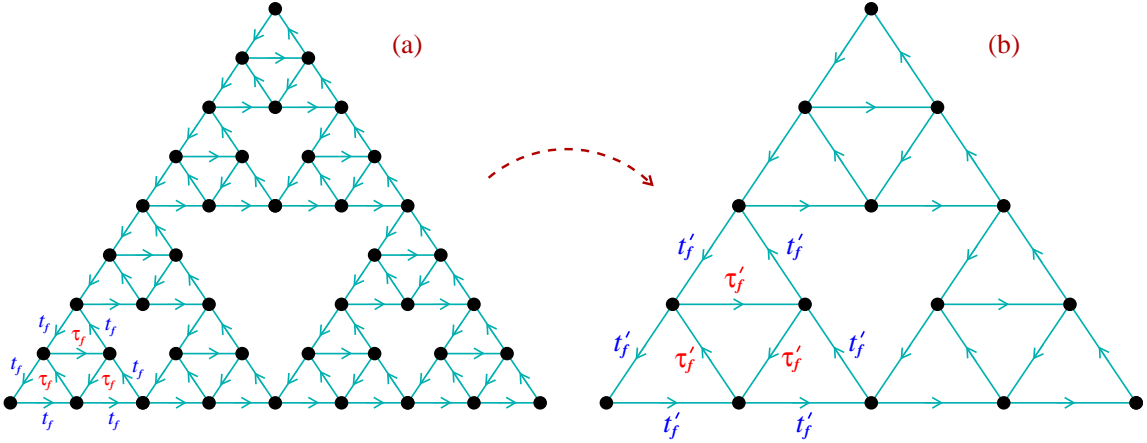}
\caption{(a) Schematic diagram of the  effective Seirpinski gasket 
fractal which is obtained by decimating the the `B' type (red) atomic 
sites from the original graphene-fractal network. The arrows show the 
direction for {\it forward} hopping from one atomic site to another atomic site 
-- for the {\it inner} triangles it is $\tau_{f}$ and for {\it outer} 
triangles it is $t_{f}$. (b) The renormalized version of (a).}
\label{lattice2}
\end{figure}

The Schr\"{o}dinger equation for the original graphene network is 
equivalently cast into a set of difference equations, viz,
\begin{equation}
(E-\epsilon_{i})\:\psi_{i}=\sum_{j}\;t_{ij} e^{i \theta_{ij}}\:\psi_{j}
\label{difference}
\end{equation}
which is utilized in decimating out a suitable subset of atoms to reduce 
the infinite graphene-fractal to a standard triangular SPG. In this 
triangular SPG, there is a uniform value of the effective on-site 
potential $\epsilon$, which, of course, is a function of energy now. The 
nearest neighbor hopping integrals assume two different values depending 
on the phase associated with them. The distribution is illustrated in 
Fig.~\ref{lattice2}(a). The bonds of the {\it inner} triangles are 
associated with the hopping integrals $\tau_{f}$ (forward) and $\tau_{b}$ 
(backward), while those along the {\it outer} triangle are designated by 
$t_{f}$ and $t_{b}$ respectively. The values of these parameters are 
given by,  
\begin{eqnarray}
\epsilon &=& \epsilon_{0}+\dfrac{2t_{0}^{2}}
{E-\epsilon_{0}}\nonumber\\
\tau_{f} &=& \dfrac{t_{0}^{2}e^{2i\theta}}
{E-\epsilon_{0}},\ 
\tau_{b} = \dfrac{t_{0}^{2}e^{-2i\theta}}
{E-\epsilon_{0}}\nonumber\\
t_{f}  &=&  \dfrac{t_{0}^{2}e^{-i\theta}}
{E-\epsilon_{0}},\
t_{b}  =  \dfrac{t_{0}^{2}e^{i\theta}}
{E-\epsilon_{0}}
\label{hsg}
\end{eqnarray}

The effective SPG network (Fig.~\ref{lattice2}(a)) is easily renormalized 
(Fig.~\ref{lattice2}(b)) and the renormalized values of the parameters 
are, 
\begin{eqnarray}
\epsilon^{\prime} &=& \epsilon + 
2(ut_{f}+u^{*}t_{b})\nonumber\\
\tau_{f}^{\prime} &=& vt_{f}+w^{*}t_{b},\ 
\tau_{b}^{\prime} = wt_{f}+v^{*}t_{b}\nonumber\\
t_{f}^{\prime} &=& vt_{f}+w^{*}t_{b},\ 
t_{b}^{\prime} = wt_{f}+v^{*}t_{b}
\label{spg}
\end{eqnarray}
\begin{eqnarray}
\text{where,\quad} u &=&(p+rp^{*})/\mathcal{D},\ 
v =(q+rs^{*})/\mathcal{D},\ 
w =(s+rq^{*})/\mathcal{D}\nonumber\\
u^{*}&=&(p^{*}+r^{*}p)/\mathcal{D},\ v^{*}=(q^{*}+r^{*}s)/\mathcal{D},\ 
w^{*}=(s^{*}+r^{*}q)/\mathcal{D}\nonumber\\
\mathcal{D}&=& 1-rr^{*}\nonumber\\
\nonumber\\
\text{Here,\quad}
p &=&\dfrac{(E-\epsilon)t_{b}}{\delta_{0}},\ 
q =\dfrac{(E-\epsilon)t_{f}+\tau_{b}t_{b}}
{\delta_{0}}\nonumber\\
p ^{*}&=&\dfrac{(E-\epsilon)t_{f}}{\delta_{0}},\ 
q ^{*}=\dfrac{(E-\epsilon)t_{b}+\tau_{f}t_{f}}
{\delta_{0}}\nonumber\\
r &=&\dfrac{(E-\epsilon)\tau_{f}+(\tau_{b})^{2}}
{\delta_{0}},\ 
s =\dfrac{\tau_{b}t_{f}}{\delta_{0}}\nonumber\\
r ^{*}&=&\dfrac{(E-\epsilon)\tau_{b}+(\tau_{f})^{2}}
{\delta_{0}},\ 
s^{*}=\dfrac{\tau_{f}t_{b}}{\delta_{0}}\nonumber\\
\delta_{0}&=&(E-\epsilon)^{2}-\tau_{f}\tau_{b}\nonumber
\end{eqnarray}

The above set of recursion relations will now be analyzed to see the 
effects of a fractal distribution of holes in a graphene sheet with  
magnetic flux trapped in a selected set of the hexagons.  
\section*{3. Results and discussion}

\subsection*{3.1 Local Density of States (LDOS) at the bulk atomic sites}

We have calculated the LDOS at the `A' type (black) bulk atomic sites for 
an infinite graphene-fractal using the Green's function and a standard 
real space decimation technique~\cite{southern}. The LDOS is given by,
\begin{equation}
\rho_{00}(E)= \lim_{\eta \rightarrow 0}\,\left[-\dfrac{1}{\pi}\,
\text{Im}\,G_{00}(E+i\eta)\right]
\end{equation}
where, $G_{00}(E)$ is the local Green's function at the `A' type (black) 
bulk atomic sites. To obtain $G_{00}$, the recursion relations 
Eq.~\eqref{spg} are iterated with a small imaginary part added to the 
energy $E$ until the magnitude of the nearest neighbor hopping goes to 
zero (or, equivalently, becomes less than a small pre-assigned quantity). 
The on-site potential flows to a fixed point value $\epsilon^*$, and in 
this limit $G_{00} = (E + i\eta - \epsilon^*)^{-1}$.

\subsubsection*{3.1.1 The zero flux case}

\begin{figure}[ht]
\centering 
\includegraphics[clip,width=13cm,angle=-90]{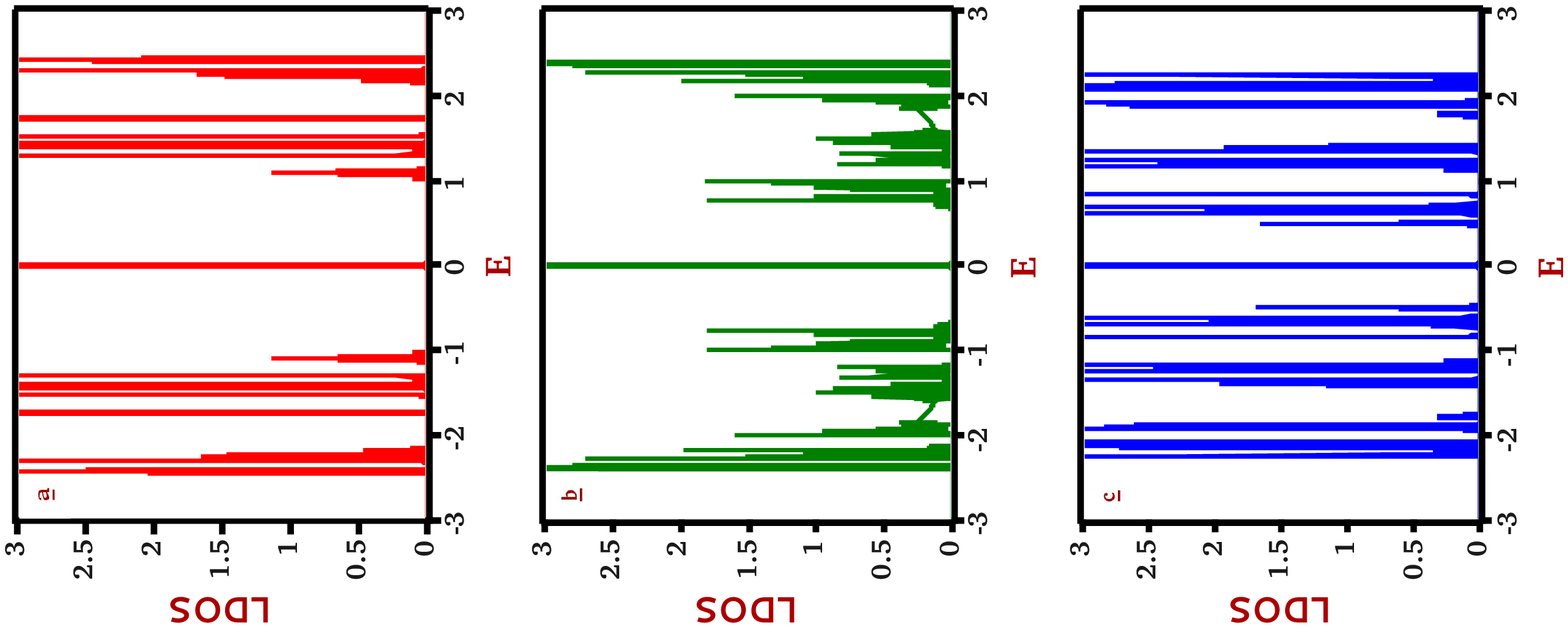}
\caption{Plots for Local Density of states (LDOS) with the energy of the 
electron ($E$) at the bulk atomic sites (`A' type) of an infinite 
graphene-fractal network for different values of magnetic flux $\Phi$. 
(a) represents the case for $\Phi=0$, (b) represents the case for 
$\Phi=\Phi_{0}/4$ and (c) represents the case for $\Phi=\Phi_{0}/2$. We 
have set $\epsilon_{0}=0$ and $t_{0}=1$.}
\label{ldos1}
\end{figure}

In the absence of any magnetic flux $\Phi$, the LDOS shows a fragmented 
structure (Fig.~\ref{ldos1}, top panel). Most of these fragmented states 
are localized states, which can be verified by studying the flow the 
hopping integral under successive iterations, keeping the energy at a 
particular value. With any arbitrarily chosen energy $E$ at which the 
LDOS is non-zero, the hopping integral flows to zero under RSRG steps.

Interestingly, such a graphene-fractal is found to sustain an infinite 
number of extended eigenstates. These eigenstates coexist with  the 
localized states. This turns out to be a generic feature of the 
underlying triangular Sierpinski geometry~\cite{wang1,arunava1}. To 
extract such extended states we need to set the energy of the electron 
$E=\epsilon^{(\ell)}$, where, $\epsilon^{(\ell)}$ represents the 
effective on-site potential at a bulk atomic site after the $\ell$-th 
stage of renormalization. A look at the recursion relations given in 
Eq.~\eqref{spg} with $\Phi = 0$ will reveal that, setting 
$E=\epsilon^{(\ell)}$ immediately leads to a {\it `two-cycle'} fixed 
point, viz, $\epsilon^{(\ell+2)} = \epsilon^{(\ell+1)} = 
\epsilon^{(\ell)}$ and $t^{(\ell+2)} = -t^{(\ell+1)} = t^{(\ell)}$ 
beginning at a certain stage $\ell$ of renormalization. $E - 
\epsilon^{\ell} = 0$ gives a polynomial equation of $E$, the real 
solutions of which will yield the values of energies at which the states 
are extended. Of course, the solutions will have to lie within the 
spectrum of the graphene-fractal.

For example, if we set $E = \epsilon$ at the basic stage $\ell = 0$, with 
$\epsilon_0 = 0$, and $t_0 = 1$, then $E = \pm \sqrt{2}$ will be energies 
for two extended states sustained by the fractal lattice in the absence 
of any magnetic flux. If we set $E = \epsilon^{(1)}$ on a one step 
renormalized lattice and with zero flux, it leads to an equation,
\begin{equation}
E^6 - 8 E^4 + 17 E^2 - 10 = 0
\end{equation}
The roots are, $ E = \pm 1,\ \pm \sqrt{2},\ \pm \sqrt{5}$. Out of these 
$E = \pm 1$ are `spurious' roots, not included in the spectrum of the 
graphene-fractal, while the others are there, and correspond to two-cycle 
fixed points of the RSRG recursion relations beginning at the stage $\ell 
= 1$. These are extended eigenstates. It is to be noted that the roots 
evolving from the solution of the equation $E = \epsilon^{(\ell)}$ are 
also included in the spectrum of $E = \epsilon^{(\ell+1)}$. This is again 
a generic feature of the SPG~\cite{arunava1}.

\subsubsection*{3.1.2 The non-zero flux cases}
\begin{figure}[ht]
\centering 
\includegraphics[clip,width=5.4in,angle=-90]{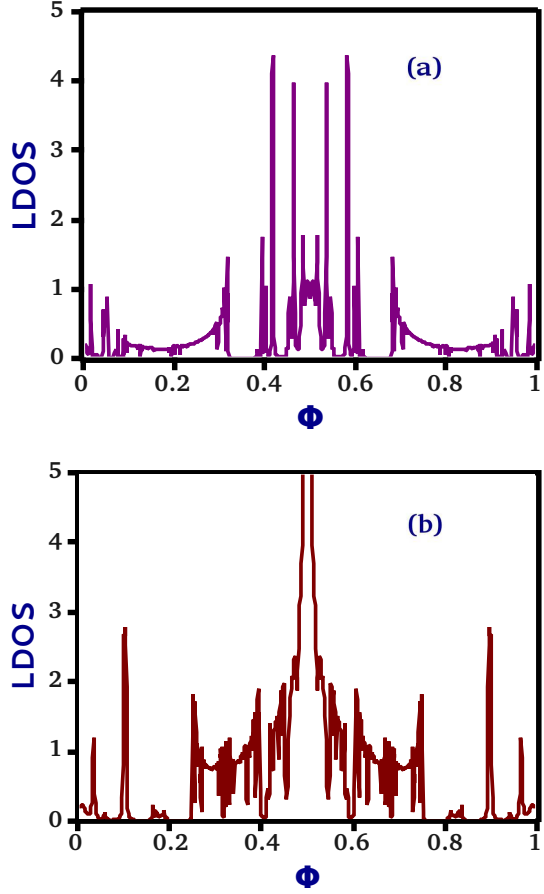}
\caption{Variation of Local Density of States (LDOS) with the magnetic 
flux $\Phi$ at the bulk atomic sites (`A' type) of an infinite graphene-
fractal network. (a) for energy $E=\sqrt{2}$ and (b) for energy 
$E=\sqrt{5}$. We have set $\epsilon_{0}=0$ and $t_{0}=1$.}
\label{ldos2}
\end{figure}

\vskip .2in
{\textbf{\itshape (a) The general character of the spectrum:}}
\vskip .1in
\noindent
As we set the magnetic flux $\Phi$ in each hexagon to a non-zero value, 
continuous bands are seen to be created in the LDOS profile 
(Fig.~\ref{ldos1}, middle panel). In this we have shown the LDOS for 
$\Phi=\Phi_{0}/4$ and it clearly shows the continua in the range $1 < |E| 
< 2$. In such continua the states are {\it `extended'}, as has been 
examined by picking up any energy in this domain arbitrarily, and 
studying the flow of the hopping integral. The hopping integral in 
general, oscillates chaotically and remain non-zero for an indefinite 
number of iterations. The LDOS profile shows a continuum zone for all 
non-zero values of $0 < \Phi < \Phi_{0}/2$. Energy eigenvalues, picked up 
arbitrarily from any such continuous zone, corresponds to an extended 
eigenstate of the system. 
The localized eigenstates are also there. The spectrum is a mixture of 
the bands of extended states and clusters of localized states. However, 
it is not apparent whether there is any possibility of a metal-insulator 
transition driven by the magnetic flux.
At $\Phi=\Phi_{0}/2$ the fragmented character 
in the LDOS profile is restored again (Fig.~\ref{ldos1}, bottom panel). 
\vskip .1in
\noindent
{\textbf{\itshape (b) Fixed points:}}
\vskip .1in
\noindent
In the presence of a magnetic flux, the extraction of a fixed point of 
the transformations Eq.~\eqref{spg} is a non-trivial issue. Nevertheless, 
it is possible to work out the case for $\ell = 1$. Setting $E = 
\epsilon^{(1)}$ leads to the equation, 
\begin{equation}
{\mathit f}(E)= \cos 6\theta
\label{onestep}
\end{equation}
where, ${\mathit f}(E)=(E^6 - 8 E^4 + 17 E^2 - 8)/2$ and $\theta = 2 \pi 
\Phi/6 \Phi_0$. 
\begin{figure}[ht]
\centering 
\includegraphics[clip,width=3.8in]{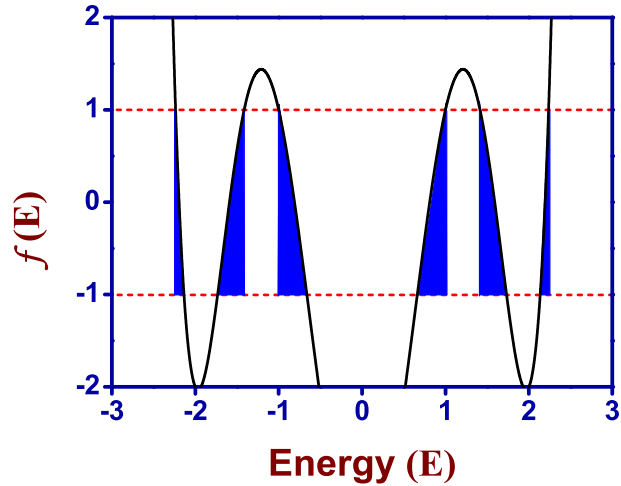}
\caption{Plot of energy function ${\mathit f}(E)$ on the LHS of 
Eq.~\eqref{onestep} with energy $E$. The shaded portions under the curve 
shows the values of energy for which the value of ${\mathit f}(E)$ lies 
within $\pm1$.}
\label{engfunc}
\end{figure}
Clearly, for any energy $E$ lying within the spectrum, 
whenever the left hand side of Eq.~\eqref{onestep} remains bounded by $\pm 
1$, one can tune the magnetic flux appropriately to satisfy the equation 
$E = \epsilon^{(1)}$. For this special value of the magnetic flux, a two-
cycle fixed point behavior is obtained. We depict the variation of the 
left hand side of Eq.~\eqref{onestep} in Fig.~\ref{engfunc}. In principle, 
any energy within the shaded zone (provided it is included in the overall 
energy spectrum) can be made to correspond to an extended eigenfunction 
by a magnetic flux, whose value can be estimated from Eq.~\eqref{onestep}. 
For example, if we set the Fermi level at $E = 0.7$ (in unit of $t_0$), 
then it can be worked out to find that a value of the magnetic flux $\Phi 
= 0.38178 \Phi_0$ throws the parameter space 
$(\epsilon^{(\ell)},t^{(\ell)})$ into a two-cycle fixed point beginning 
at the $\ell = 1$. As soon as one shifts away from this special value 
of the flux, the hopping integrals usually flow to zero as the 
renormalization progresses. This implies that, once we fix the Fermi 
level at a special value, the entire graphene-fractal can be made to 
conduct or act as an insulator by tuning the magnetic flux piercing the 
elementary hexagons.
\vskip .1in
\noindent
{\textbf{\itshape (c) Flux controlled behavior of the eigenstates:}}
\vskip .1in
\noindent
Certain points are of interest, and need to be appreciated in the context 
of flux controlled behavior of the eigenstates in a graphene-fractal. To 
appreciate this discussion we first illustrate in Fig.~\ref{ldos2} the 
variation of the LDOS against the magnetic flux at special values of 
energy $E = \sqrt{2}$ (Fig.~\ref{ldos2}(a)), and for $E = \sqrt{5}$ 
(Fig.~\ref{ldos2}(b)). In each diagram it is clear that the selected 
energy can be brought inside or thrown out of the energy spectrum by 
tuning the magnetic flux. We specially focus in and around $\Phi = 
\Phi_0/2$.

Let us first consider the case when $E = \sqrt{2}$. There is clearly a  
continuous distribution of flux values in the neighborhood of $\Phi = 
\Phi_0/2$, viz, for $0.495\Phi_0 < \Phi <0.505\Phi_0$, for which the LDOS 
is non-zero. For this range of magnetic flux, with the selected energy, 
the hopping integral does not flow to zero, but becomes small and keeps 
on oscillating around a value $\sim 10^{-2}$. This implies that the 
wavefunction really does not decay, but the overlap of amplitudes at 
distant lattice sites becomes very small. This tempts us to categorize 
such states as {\it critical}~\cite{domany}, or at least they have very 
large localization lengths.

Choosing $E = \sqrt{5}$ leads to a more interesting scenario. We now have 
a sharply localized eigenstate for $\Phi = \Phi_0/2$. The LDOS is 
isolated at $\Phi = \Phi_0/2$, and is highly degenerate. Setting $E = 
\sqrt{5}$, and $\Phi = \Phi_0/2$ makes the hopping matrix element flow to 
zero quickly. This observation speaks in favor of the localized character 
of the states. Interestingly, in the immediate neighborhood of the center 
of the spectrum ($\Phi = \Phi_0/2$), the hopping integral displays non-
zero values under successive iterations, indicating either completely 
extended states, or critical ones (in the sense as discussed above). We 
thus encounter a possibility of a {\it reentrant crossover} in the nature 
of the spatial extension of the wavefunction at $E = \sqrt{5}$ by tuning 
the magnetic flux around the half flux quantum. 

\subsection*{3.2 Transmission characteristics of the finite graphene-
fractal network}

\begin{figure}[ht]
\centering 
\includegraphics[clip,width=12cm]{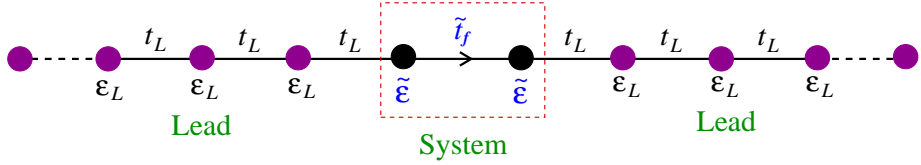}
\caption{Schematic diagram of the effective dimer clamped between two 
semi-infinite ordered leads.}
\label{dimer}
\end{figure}

To get the two terminal end-to-end transmission coefficient for a finite 
sized graphene-fractal, we clamp the system between two semi-infinite 
ordered leads. The leads, in the tight binding model, are described by a 
constant on-site potential $\epsilon_{L}$ and a nearest neighbor hopping 
integral $t_{L}$. We then successively renormalize the system to reduce 
it into an effective dimer (Fig.~\ref{dimer}) consisting of two 
`renormalized' atoms, each having an effective on-site potential equal to 
$\tilde{\epsilon}$ and with an effective hopping integral $\tilde{t}$. 

The transmission coefficient across the effective dimer is given 
by~\cite{stone}
\begin{equation}
T=\dfrac{4\sin^{2}ka}{\left[(M_{12}-M_{21})+(M_{11}-M_{22})\cos ka 
\right]^{2}+\left[(M_{11}+M_{22})\sin ka \right]^{2}}
\end{equation}
where, 
\begin{equation*}
M_{11} =\dfrac{(E-\tilde{\epsilon})^{2}}{\tilde{t_{f}}t_{L}}
-\dfrac{\tilde{t_{b}}}{t_{L}},\ 
M_{12} =-\dfrac{(E-\tilde{\epsilon})}{\tilde{t_{f}}},\ 
M_{21} =-M_{12},\ 
M_{22} =-\dfrac{t_{L}}{\tilde{t_{f}}}
\end{equation*}
and `$a$' is the lattice constant and is taken to be equal to unity 
throughout the calculation.

\begin{figure}[ht]
\centering 
\includegraphics[clip,width=13cm,angle=-90]{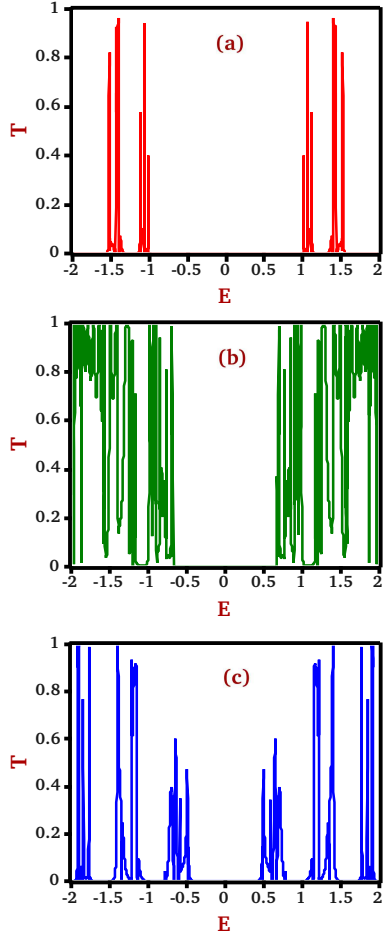}
\caption{Transmission coefficient across a $4$-th generation graphene-
fractal network for different values of magnetic flux $\Phi$. (a) 
represents the case for $\Phi=0$, (b) represents the case for 
$\Phi=\Phi_{0}/4$ and (c) represents the case for $\Phi=\Phi_{0}/2$. We 
have set $\epsilon_{0}=\epsilon_{L}=0$ and $t_{0}= t_{L}=1$.}
\label{trans}
\end{figure}

In Fig.~\ref{trans} (a), (b), and (c) we plot the two terminal 
transmission coefficient of a $4$-th generation graphene-fractal for 
$\Phi = 0$, for $\Phi = \Phi_0/4$ and $\Phi = \Phi_0/2$ respectively. The 
figures bring out the typical fragmented spectrum for zero magnetic 
field, which then gets converted into patches of continua at $\Phi = 
\Phi_0/4$. Finally, at $\Phi = \Phi_0/2$, the fragmented character is 
restored, reflecting a poorly conducting system. These results are at par 
with our discussion about the general band structure of the graphene-
fractal system.

We have also examined the AB oscillations in the transmission spectrum at 
particular values of the electron energy $E$. The oscillations have the 
typical $\Phi_0$ periodicity observed in systems even with a single loop. 
However, we do not show this result to save space.
\section*{4. Concluding remarks}
In conclusion, we have considered an infinite graphene sheet in which 
holes have been created following a Sierpinski fractal distribution. The 
electronic spectrum of the system, investigated within a tight binding 
Hamiltonian for spinless, non-interacting electrons reveal a wide variety 
in the nature of single particle states. The magnetic field piercing each 
basic hexagonal plaquette is shown to lead to absolutely continuous parts 
in the spectrum. The results are reflected in the flow of the hopping 
integrals, as observed within a real space renormalization group 
formalism. A crossover in the behavior of the wavefunction at particular 
values of the energy can be made to happen by tuning the magnetic flux. 
The graphene-fractal system is thus a candidate to be inspected more 
carefully as a potential electronic device.

\subsection*{Acknowledgment}
The first author Biplab Pal acknowledges a scholarship (IF110078) from 
DST, India.


\end{document}